\documentclass[pra,twocolumn,tightenlines,showpacs,nofootinbib,floatfix]{revtex4}

\usepackage{multirow}
\usepackage{amsmath,amsfonts,amssymb}
\usepackage{bm}
\usepackage{dcolumn}
\usepackage{graphicx}

\newcommand{\etal}{{\it et al.}}

\newcommand{\eref}[1]{Eq.~(\ref{#1})}
\newcommand{\tref}[1]{Table~\ref{#1}}

\begin{document}

\title{A development of the CI + all-order method and application to the parity-nonconserving amplitude and other properties of Pb}

\author{S.~G.~Porsev$^{1,2}$}
\author{M.~G.~Kozlov$^{2,3}$}
\author{M.~S.~Safronova$^{1,4}$}
\author{I.~I.~Tupitsyn$^{5}$}
\affiliation{ $^1$Department of Physics and Astronomy, University of Delaware, Newark, Delaware 19716, USA\\
$^2$Petersburg Nuclear Physics Institute, Gatchina, Leningrad District, 188300, Russia \\
$^3$St.\ Petersburg Electrotechnical University ``LETI'', St.\ Petersburg, Russia \\
$^4$Joint Quantum Institute, National Institute of Standards and Technology and the University of Maryland,
College Park, Maryland, 20742\\
$^5$Department of Physics, St. Petersburg State University, Ulianovskaya 1, Petrodvorets, St.Petersburg, 198504, Russia}

\date{\today}

\begin{abstract}
We have further developed and extended a method for calculation of atomic properties based
on a combination of the configuration interaction and
coupled-cluster approach. We have applied  this approach to the calculation of
different properties of atomic lead, including the energy levels,
hyperfine structure constants, electric-dipole transition amplitudes, and $E1$
parity nonconserving (PNC) amplitude for the $6p^2\,\,^3\!P_0 -
6p^2\,\,^3\!P_1$ transition. The uncertainty of the $E1$ PNC amplitude
was reduced by a factor of two in comparison with the previous most accurate
calculation [V.~A.~Dzuba \etal,\ Europhys. Lett. {\bf 7}, 413 (1988)]. Our
value for the weak charge $Q_W=-117(5)$ is in agreement with the standard model prediction.
\end{abstract}
\maketitle

\section{Introduction}
  Accurate calculations of atomic  properties of heavy atoms with several valence
electrons is a  difficult endeavor. A standard approach in atomic calculations is to separate the atomic electrons
into two groups, core and valence electrons. Then, various methods  exist to
treat  core-core, core-valence and valence-valence correlations. In particular, the valence-valence
correlations can be treated in the framework of multi-configuration Hartree-Fock (MCHF)~\cite{FisTacGai07,JonFisGod99},
relativistic multi-configuration Dirac-Fock (MCDF)~\cite{Grant,IndDes90},
or configuration interaction (CI)~\cite{KotTup87,JonYnnFis96,PorKozRei09} methods.
The core-core and core-valence correlations can be taken into account using many-body perturbation theory or
coupled-cluster method~\cite{LinMor86,Joh07,CarPalPit10}.

Complexity of calculations rapidly increases with increasing number of valence electrons.
Most precise calculations of different properties were carried out for monovalent atoms.
In particular, parity nonconserving (PNC) amplitude for the $6s-7s$ transition in atomic Cs was calculated
with uncertainty a few tenths of a percent~\cite{PorBelDer09,PorBelDer10,DzuBerFla12}, while the uncertainty of the
PNC amplitude in the $(6s^2 6p)\, ^2\!P^o_{1/2} - (6s^2 6p)\,^2\!P^o_{3/2}$ transition of three-valence Tl is an order of 
magnitude larger~\cite{DzuFlaSil87J,HarLinMar90,KozPorJoh01}.

The ground state electronic configuration of Pb atom is
$[{\rm Xe}] 4f^{14} 5d^{10} 6s^2 6p^2$. Measurements of parity nonconserving
optical rotation near the 1.279 $\mu$m, $6p^2\, ^3\!P_0 \rightarrow 6p^2\, ^3\!P_1$ magnetic dipole transition in Pb
were carried out almost 20 years ago by
Seattle~\cite{MeeVetMaj93,MeeVetMaj95} and Oxford~\cite{EdwPhiBai95}
groups, giving the ratio, $R$, of the $E1$ PNC to the $M1$
transition amplitude to be $(-9.86 \pm 0.12) \times 10^{-8}$ and
$(-9.80 \pm 0.33) \times 10^{-8}$, correspondingly. Thus, the
Seattle group achieved the experimental precision of $1.2$\%.

The quantity $R$ is proportional to $Q_W$, where the nuclear weak charge $Q_W$ at tree
level is given by the formula
\begin{equation}
Q_W \approx -N + Z\,(1-4\,{\rm sin}^2\theta_W), \label{QW}
\end{equation}
where $N$ is the number of neutrons, $Z$ is the nuclear charge, and
$\theta_W$ is the Weinberg angle. For $^{208}$Pb, this expression
gives $Q_W \approx -120$. A more accurate standard model (SM) value,
which includes radiative corrections, is
$Q_W^\mathrm{SM}=-118.79(5)$~\cite{PDG12}.

Atomic parity violation studies test the standard model of the
electroweak interaction by comparing the SM value of the weak charge
with the $Q_W$ extracted from the experiments. Such extraction requires
 an accurate calculation of the
quantity $R$. Due to complicated electronic structure of
Pb, there were only a few calculations of the PNC amplitude in the
$6p^2\, ^3\!P_0 \rightarrow 6p^2\, ^3\!P_1$
transition~\cite{NovSusKhr76,BotBlaMar87,DzuFlaSil87E}. Most accurate result for $R$ was obtained
in~\cite{DzuFlaSil87E}, where this quantity was determined with 8\%
uncertainty.

In this work we continue a development of the high-precision relativistic atomic method which
combines the configuration interaction and coupled-cluster (CI+all-order)
approaches~\cite{Koz04,SafKozJoh09}.
Initial variant of the method required either to
treat Pb as a system with two valence electrons and use $V^{N-2}$
potential, or as a four-electron system using $V^{N-4}$ potential.

In the present work, we extend the CI+all-order method to the case when
initial approximation does not correspond to the self-consistent
field of the core.
New variant of the method allowed us to consider Pb as a system with
four valence electrons but use $V^{N-2}$ potential. This requires to add
a number of so-called ``subtraction'' terms into the all-order equations.
We report both $V^{N-2}$ and $V^{N-4}$ calculations of Pb properties and conclude that
the former is more reliable and accurate. We calculated the quantity
$R$ to be $R = 10.6(4) \times 10^{-8}\,i(-Q_W/N)$, reducing its
uncertainty by a factor of 2 in comparison with~\cite{DzuFlaSil87E}.

The paper is organized as follows. In Section~\ref{MonAt} we
describe main features of our method and discuss a choice of
initial approximation. In Sections~\ref{ManyAt} and \ref{Ang} we
show how the equations, describing core-valence correlations, are modified for
a more flexible choice of the potential. In Sections~\ref{VN-2} and \ref{VN-4} we
discuss and compare the results obtained in $V^{N-2}$ and $V^{N-4}$
approximations. The last section contains concluding remarks and
acknowledgements. If not stated otherwise, atomic units ($\hbar =
|e| = m_e =1$) are used throughout.

\section{A choice of initial approximation}
\label{MonAt}
Using second quantization, the relativistic no-pair Hamiltonian
$H$  can be written as $H=H_{0}+V$~\cite{BroRav51,Johnson07}:
\begin{equation}
H_{0}=\underset{i}{\sum }\varepsilon _{i}\{a_{i}^{\dagger }a_{i}\},
\end{equation}%
\begin{eqnarray}
V &=& \frac{1}{2}\underset{ijkl}{\sum }g_{ijkl}\{a_{i}^{\dagger}a_{j}^{\dagger }a_{l}a_{k}\}
\nonumber \\
&+& \underset{ij}{\sum }(V_{DF}-U_{DF})_{ij}\{a_{i}^{\dagger }a_{j}\}.
\label{V}
\end{eqnarray}%
Here, $a_{i}^{\dagger}$ and $a_{i}$ are the creation and
annihilation operators, respectively; $\left\{ ...\right\} $
designates normal form of operators in respect to the core state
$\Psi_c$: $\left\{...\right\}|\Psi_c\rangle=0$, and the indexes $i,j,k$, and $l$ range over
\textit{all} possible single-electron states. $H_0=T+U_{DF}$ is the
Hartree-Fock-Dirac (HFD) operator for $N_{DF}$ electrons, forming
closed sub-shells. In this work we do not consider a more general
case, when $H_0$ is not a HFD operator.

Two-particle matrix elements (MEs), designated as $g_{ijkl}$, are given by
\begin{equation*}
g_{ijkl}=\int dr^{3} dr^{\prime 3}\psi _{i}^{\dagger }(\mathbf{r})\psi
_{j}^{\dagger }(\mathbf{r}^{\prime })\frac{1}{\left\vert \mathbf{r}-\mathbf{r%
}^{\prime }\right\vert }\psi _{k}(\mathbf{r})\psi _{l}(\mathbf{r}^{\prime }) ,
\end{equation*}%
where $\psi_i$ are the single-electron wave functions and $V_{DF}$
is the frozen-core Dirac-Fock (DF) potential determined as
\begin{equation}
(V_{DF})_{ij}=\overset{N_{c}}{\underset{b=1}{\sum }}\left(
g_{ibjb}-g_{ibbj}\right) \equiv \overset{N_{c}}{\underset{b=1}{\sum }}
\tilde{g}_{ibjb},
\label{VDF}
\end{equation}%
with $N_{c}$ being the number of the core electrons.

There is certain flexibility in choosing potential $U_{DF}$, which
defines initial approximation $H_0$ and enters~\eref{V}. It is
convenient to determine $U_{DF}$ as DF potential for $N_{DF}$
electrons:
\begin{equation}
(U_{DF})_{ij}\equiv \overset{N_{DF}}{\underset{b=1}{\sum
}}\tilde{g}_{ibjb}. \label{UDF}
\end{equation}%
For monovalent atoms the natural choice is $N_{DF}=N_c=N-1$, which
leads to $U_{DF}=V_{DF}$. This is often referred to as $V^{N-1}$
approximation. With such a choice the second term in Eq.~(\ref{V})
vanishes.

For multivalent atoms it is sometimes convenient to choose $N_{DF}>N_{c}$.
The dominant configuration of Pb ground state is $[\dots]\,6s^2 6p^2$, thus Pb can be considered as an atom with 4 valence
electrons. However, we can include two $6s$ electrons, forming closed sub-shell,
in the initial HFD self-consistency procedure and
construct the basis set in $V^{N-2}$ approximation. Then the number
of the core electrons is $N_c=N-4$, while $N_{DF}=N-2 > N_c$. As a
result, there will be only partial cancellation of the terms
determined by the potential $V_{DF}$ and the Dirac-Fock field
$U_{DF}$ in Eq.~(\ref{V}).

We designate the difference between $V_{DF}$ and $U_{DF}$ as $U$.
Then, for a single-electron matrix element:
\begin{equation}
U_{ij} = (U_{DF}-V_{DF})_{ij} .
\end{equation}%
Performing angular reduction yields
$$U_{ij}=\delta_{\varkappa _{i}\varkappa _{j}}\delta _{m_{i}m_{j}}U(ij),$$
where
\begin{eqnarray}
U(ij) &=& \delta _{\varkappa _{i}\varkappa _{j}}\overset{N_{DF}}
{\underset{b=N_c+1}{\sum}} \left[ \sqrt{\frac{2j_{b}+1}{2j_{i}+1}}X_{0}\left(
ibjb\right) \right. \nonumber \\
&+& \left. \underset{L}{\sum }\frac{(-1)^{j_{i}+j_{b}+L}}{\left(
2j_{i}+1\right) }X_{L}\left( bijb\right) \right] .
\label{Uij}
\end{eqnarray}%
Here, the sum over index $b$ means the sums over principal quantum number $n_b$ and relativistic quantum number
$\varkappa_b=(l_b-j_b)(2j_b+1)$, where $l_b$ and $j_b$ are the orbital and total angular momenta.
We use notation%
\begin{eqnarray}
&& X_{L}\left( mnab\right) = \nonumber \\
&&\left( -1\right) ^{L}\langle \varkappa_{m}||C^{L}||\varkappa _{a}\rangle
\langle \varkappa _{n}||C^{L}||\varkappa_{b}\rangle R_{L}(mnab),
\end{eqnarray}%
where $R_{L}(mnab)$ is relativistic Slater integral and
$\langle \varkappa_m||C^{L}||\varkappa_a\rangle $ is the reduced matrix element of a
normalized spherical harmonic given by
\begin{eqnarray}
&&\langle \varkappa_m ||C^L|| \varkappa_a \rangle = \xi(l_m+l_a+1) (-1)^{j_m+1/2}  \nonumber  \\
&& \times \sqrt{(2j_m+1)(2j_a+1)}
\left(
\begin{array}{ccc}
 j_m & j_a & L \\
-1/2 & 1/2 & 0
\end{array}
\right),
\end{eqnarray}
where
\begin{center}
$ \xi(x) =
\left\lbrace
\begin{array}{l}
1,\, {\rm if}\,\, x\,\, {\rm is\,\, even}  \\
0,\, {\rm if}\,\, x\,\, {\rm is\,\, odd}
\end{array}
\right.$ .
\end{center}

\section{Effective Hamiltonian for multivalent atoms}
\label{ManyAt}

The wave functions and energy levels of the valence electrons can be found
by solving the multiparticle relativistic equation~\cite{DzuFlaKoz96}:
\begin{equation}
H_{\rm eff}(E_n) \Phi_n = E_n \Phi_n,
\label{Heff}
\end{equation}
where the effective Hamiltonian is defined as
\begin{equation}
H_{\rm eff}(E) = H_{\rm FC} + \Sigma(E),
\label{Heff1}
\end{equation}
with $H_{\rm FC}$ being the Hamiltonian in the frozen-core approximation.
The energy-dependent operator $\Sigma(E)$ accounts for virtual excitations of the core electrons.
It is constructed using the second order many-body perturbation theory in the CI+MBPT approach \cite{DzuFlaKoz96} or linearized
coupled cluster single-double (LCCSD) method in the CI+all-order approach~\cite{SafKozJoh09}.

It is convenient to decompose the effective Hamiltonian $H_{\rm eff}(E)$ into two parts $H_{\rm eff}(E) = H_1+H_2$,
where $H_1$ represents the one-body part of the Hamiltonian and $H_2$ represents the two-body part of the Coulomb interaction.
In this work, we disregard the Breit interaction as well as three-electron part of the effective Hamiltonian \cite{DzuFlaKoz96}.

The energy-dependent operator $\Sigma$ is also separated into two parts,  $\Sigma = \Sigma_1 + \Sigma_2$, where $\Sigma_1$ and $\Sigma_2$
describe one- and two-body parts of core-valence correlations, respectively. The expressions for single-electron matrix elements
of these operators, $(\Sigma_1)_{ij}$ and
$(\Sigma_2)_{ijkl}$, obtained in the LCCSD method approximation for the case
$U=0$ (i.e., when $U_{DF}=V_{DF}$), are given and discussed in detail in Ref.~\cite{SafKozJoh09}.

In the case of  $U \neq 0$, we should add the terms linear in $U$ to the equations
for the cluster amplitudes calculated in the framework of the LCCSD approximation.
The resulting LCCSD equations derived for this more general case are presented below:
\begin{widetext}
\begin{subequations}
\label{Sigma}
 \begin{eqnarray}
\Sigma _{ma} &=&\mathrm{LCCSD}\text{ }-U_{ma}+\underset{b}{\sum }\rho
_{mb}U_{ba}-\underset{n}{\sum }\rho _{na}U_{mn}-\underset{bn}{\sum }\tilde{%
\rho}_{mnab}U_{bn},
\\
\Sigma _{mnab} &=&\mathrm{LCCSD}\text{ }-\underset{r}{\sum }\tilde{\rho}%
_{mrab}U_{nr}+\underset{c}{\sum }\tilde{\rho}_{mnac}U_{cb},
\\
\Sigma _{mv} &=&\mathrm{LCCSD}\text{ }+\underset{b}{\sum }\frac{\Sigma
_{mb}U_{bv}}{\tilde{\varepsilon}_{v}-\varepsilon _{v}+\varepsilon
_{b}-\varepsilon _{m}}-\underset{bn}{\sum }\frac{\tilde{\Sigma}_{mnvb}U_{bn}%
}{\tilde{\varepsilon}_{v}+\varepsilon _{b}-\varepsilon _{mn}},
\\
\Sigma _{mnva} &=&\mathrm{LCCSD}\text{ }-\underset{r}{\sum }\frac{\tilde{%
\Sigma}_{mrva}U_{nr}}{\tilde{\varepsilon}_{v}+\varepsilon _{a}-\varepsilon
_{mr}}+\underset{c}{\sum }\frac{\Sigma _{mnvc}U_{ca}}{\tilde{\varepsilon}%
_{v}+\varepsilon _{c}-\varepsilon _{mn}}
+\underset{c}{\sum }\frac{\Sigma _{nmac}U_{cv}}{\tilde{\varepsilon}%
_{v}-\varepsilon _{v}+\varepsilon _{ca}-\varepsilon _{mn}},
\\
\Sigma _{mnvw} &=&\mathrm{LCCSD}\text{ }+\underset{c}{\sum }\left( \frac{%
\Sigma _{mnvc}U_{cw}}{\tilde{\varepsilon}_{v}+\tilde{\varepsilon}%
_{w}-\varepsilon _{w}+\varepsilon _{c}-\varepsilon _{mn}}+\frac{\Sigma
_{nmwc}U_{cv}}{\tilde{\varepsilon}_{w}+\tilde{\varepsilon}_{v}-\varepsilon
_{v}+\varepsilon _{c}-\varepsilon _{mn}}\right),
\end{eqnarray}%
\end{subequations}
\end{widetext}
where $\Sigma_{ij} \equiv (\Sigma_1)_{ij}$, $\Sigma_{ijkl} \equiv (\Sigma_2)_{ijkl}$, and 
$\varepsilon_{i}$ are one-electron Dirac-Fock energies and we use notation $\varepsilon_{ij} \equiv \varepsilon_i + \varepsilon_j$.
The symbol tilde over $\varepsilon_{v,w}$ reflects the fact that the effective Hamiltonian~(\ref{Heff1}) 
is energy dependent~\cite{SafKozJoh09}. A definition of the tided energy depends on 
the choice of initial approximation and will be discussed in Sec.~\ref{VN-2}.

The terms labeled ``LCCSD'' in $\Sigma _{mv}$, $\Sigma _{mnva}$, and
$\Sigma _{mnvw}$ are given by the right hand sides of Eqs.~(22-24) in
Ref.~\cite{SafKozJoh09}. The core amplitudes $\Sigma_{ma}$ and
$\Sigma_{mnab}$ are obtained from the core coefficients $\rho_{ma}$
and $\rho_{mnab}$ (given, for example, in~\cite{BluJohLiu89}) using simple
relations
\begin{eqnarray}
\Sigma_{ma} &=& \rho_{ma} (\varepsilon_a - \varepsilon_m), \nonumber \\
\Sigma_{mnab} &=& \rho_{mnab} (\varepsilon_a + \varepsilon_b - \varepsilon_m - \varepsilon_n ).
\end{eqnarray}

It is easy to verify that the expressions for $\Sigma _{mnva}$ and $\Sigma _{mnvw}$
remain the same when we transpose the indexes $m\leftrightarrow n$ and $a\leftrightarrow
v$ (or $m\leftrightarrow n$ and $w\leftrightarrow v$), maintaining original symmetry of the all-order LCCSD
equations.
\section{Angular reduction}
\label{Ang}
Using the formulas

\begin{eqnarray*}
\Sigma_{li} &=& \delta _{\varkappa_l \varkappa_i} \delta_{m_l m_i} \Sigma(li) , \\
\Sigma_{lnib} &=& \sum_{kq} \frac{1}{\sqrt{[j_l] [j_b]}} C_{j_i m_i\, kq}^{j_l m_l} C_{kq\, j_n m_n}^{j_b m_b} \Sigma_k(lnib) \\
&=& \sum_{kq} (-1)^{j_i - j_n + m_i + m_n}
\left(
\begin{array}{ccc}
j_i & k &  j_l \\
m_i & q & -m_l%
\end{array}%
\right) \nonumber \\
&\times& 
\left(
\begin{array}{ccc}
j_b &  j_n &  k \\
m_b & -m_n & -q%
\end{array}%
\right) \Sigma_k(lnib),
\end{eqnarray*}
we performed angular reduction, arriving at
\begin{widetext}
\begin{eqnarray}
\label{sigma_ang}
\Sigma (ma) &=&\mathrm{LCCSD}-U(ma)+\delta _{\varkappa _{m}\varkappa_{b}}
\sum_{n_{b}}U(ba)\rho (mb)-\delta _{\varkappa _{n}\varkappa_{a}}\sum_{n_{n}}U(mn)\rho (na)  \nonumber \\
&-& \delta_{\varkappa_n \varkappa_b} \sum_{n_n n_b \varkappa_b}
\sqrt{\frac{\left[ j_{b}\right] }{\left[ j_{a}\right] }}U(bn)\tilde{\rho}_{0}(mnab),
\nonumber  \\
\Sigma _{k}(mnab) &=&\mathrm{LCCSD}-\delta _{\varkappa _{n}\varkappa
_{r}}\sum_{n_{r}}U(nr)\tilde{\rho}_{k}(mrab)+\delta _{\varkappa
_{c}\varkappa _{b}}\sum_{n_{c}}U(cb)\tilde{\rho}_{k}(mnac),
\nonumber  \\
\Sigma (mv) &=&\mathrm{LCCSD}+\delta _{\varkappa _{m}\varkappa
_{b}}\sum_{n_{b}}\frac{U(bv)\Sigma (mb)}{\tilde{\varepsilon}_{v}-\varepsilon
_{v}+\varepsilon _{b}-\varepsilon _{m}}-\delta _{\varkappa _{n}\varkappa
_{b}}\sum_{n_{n}n_{b}\varkappa _{b}}\sqrt{\frac{\left[ j_{b}\right] }{\left[
j_{v}\right] }}\frac{U(bn)\tilde{\Sigma}_{0}(mnvb)}{\tilde{\varepsilon}%
_{v}+\varepsilon _{b}-\varepsilon _{mn}},
 \\
\Sigma _{k}(mnvb) &=&\mathrm{LCCSD}-\delta _{\varkappa _{n}\varkappa
_{r}}\sum_{n_{r}}\frac{U(nr)\tilde{\Sigma}_{k}(mrvb)}{\tilde{\varepsilon}%
_{v}+\varepsilon _{b}-\varepsilon _{mr}}+\delta _{\varkappa _{c}\varkappa
_{b}}\sum_{n_{c}}\frac{U(cb)\Sigma _{k}(mnvc)}{\tilde{\varepsilon}%
_{v}+\varepsilon _{c}-\varepsilon _{mn}} \nonumber \nonumber  \\
&+& \delta _{\varkappa _{c}\varkappa _{v}}\sum_{n_{c}}\frac{U(cv)\Sigma
_{k}(nmbc)}{\tilde{\varepsilon}_{v}-\varepsilon _{v}+\varepsilon
_{cb}-\varepsilon _{mn}},
\nonumber  \\
\Sigma_k (mnvw) &=& \mathrm{LCCSD} + 
\delta_{\varkappa_c \varkappa_w} \sum_{n_c} \frac{U(cw) \, \Sigma _{k}(mnvc)}
{\tilde{\varepsilon}_{v}+\tilde{\varepsilon}_{w}-\varepsilon _{w}+\varepsilon_{c}-\varepsilon_{mn}} 
+ \delta _{\varkappa _{c}\varkappa _{v}}\sum_{n_{c}} \frac{U(cv) \, \Sigma_k(nmwc)}
{\tilde{\varepsilon}_w + \tilde{\varepsilon}_v - \varepsilon_v + \varepsilon_c - \varepsilon_{mn}}, \nonumber 
\end{eqnarray}%
\end{widetext}
where we use notation $[x] \equiv 2x+1$.
\section{$V^{N-2}$ approximation}
\label{VN-2}
In this section we describe a construction of the basis set and calculation
of the low-lying energy levels, hyperfine structure (HFS) constants, $E1$ transition
amplitudes, and $E1$ PNC amplitude for the $6p^2\,\,^3\!P_0 - 6p^2\,\,^3\!P_1$ transition
in $V^{N-2}$ approximation.
\subsection{ Basis set and energy levels}
The basis set was constructed in the framework of Dirac-Fock-Sturm (DFS) approach.
We start from a solution of the Dirac-Fock equations (disregarding the Breit interaction)
for the [$1s^2,...,5d^{10}, 6s^2$] closed shells:
\begin{equation}
\hat H_0\, \psi_c = \varepsilon_c \,\psi_c,
\end{equation}
where $H_0$ is the relativistic DF Hamiltonian
\cite{DzuFlaKoz96,SafKozJoh09} and $\psi_c$ and $\varepsilon_c$ are the
single-electron wave functions and energies, respectively. Note that
both $6s$ electrons were included  in the initial self-consistency
procedure.

As a next step, all orbitals up to the $6s$ were frozen and the $7,8s$,
6--8$p$, $6d$, and $4f$ orbitals were constructed in respective
$V^{N-2}$ potential. Higher virtual orbitals were obtained by solving the DFS equations described
in~\cite{TupVolGla05,TupKozSha10}. The resulting basis set includes the Dirac-Fock functions for
the occupied core and valence orbitals and the Dirac-Fock-Sturm
functions for virtual orbitals and contains six
partial waves with the orbitals up to $32s, 32p, 32d, 30f, 25g$, and
$25h$.

At the stage of CI calculation we consider Pb as a
4-valence atom. We construct the set of configurations that contains single and
double excitations of the electrons from lowest-lying
configurations ($6s^2\,6p^2$, $6s^2\,6p\, 7p$, and $6s^2\,6p\, 8p$
for even-parity states and $6s^2\,6p\,7s$, $6s^2\,6p\,6d$, and
$6s^2\,6p\,8s$ for odd-parity states) to the $7-22s$, $6-22p$,
$6-17d$, $4-16f$, and $5-8g$ orbitals. We checked that triple
excitations from the low-lying configurations only slightly
change the energy levels. Then, we solved the multiparticle
relativistic Schr\"{o}dinger equation
for four valence electrons to find the eigenvectors and eigenvalues
for the low-lying states.

\begin{table*}[ht]
\caption{ $V^{N-2}$ approximation. Theoretical and
experimental~\cite{RalKraRea11} energy levels of Pb (in cm$^{-1}$).
Four-electron binding energies are given in the first row for the
ground state, energies in other rows are counted from the ground
state. Experimental binding energy of the ground state is calculated
as a sum of 4 ionization potentials (IPs):
IP(Pb$^+$)+IP(Pb$^{2+}$)+IP(Pb$^{3+}$)+IP(Pb$^{4+}$). Results of the
CI, CI+MBPT, and CI+all-order calculations are given in columns
labeled ``CI'', ``CI+MBPT'', and ``CI+All''. Corresponding relative
differences of these three calculations with the experiment are
given in percentages. In the 2nd column the electronic terms from the
NIST database~\cite{RalKraRea11} are listed. In the 3rd column the
electronic terms obtained in this calculation are given, when they differ from
the NIST's ones. In the columns 4-6 we give the Land\'e $g$ factors
for the present calculation, $LS$-coupling scheme, and the experiment.}
\label{EnVN-2}%
\begin{ruledtabular}
\begin{tabular}{rcccccrrrrrcc}
\multicolumn{1}{r}{\multirow{2}{*}{Conf}} &
\multicolumn{2}{c}{\multirow{1}{*}{Term}} &
\multicolumn{3}{c}{\multirow{1}{*}{$g$ factor}} &
\multicolumn{1}{c}{\multirow{2}{*}{CI}} &
\multicolumn{1}{l}{\multirow{2}{*}{CI+MBPT}} &
\multicolumn{1}{l}{\multirow{2}{*}{CI+All}} &
\multicolumn{1}{l}{\multirow{2}{*}{Exper.}} &
\multicolumn{3}{c}{Differences (\%)} \\
\cline{11-13}
\cline{4-6}
&\multicolumn{1}{c}{NIST}&\multicolumn{1}{c}{Present}
&\multicolumn{1}{c}{(calc.)}
&\multicolumn{1}{c}{$LS$-coupling}
&\multicolumn{1}{c}{(exp.)}
&&&&& \multicolumn{1}{r}{CI} & \multicolumn{1}{r}{CI+MBPT} & \multicolumn{1}{c}{CI+All} \\
\hline \\
$6p^2$ & $^3\!P_0$   &          &      &      &      & 756855 & 780823 & 782396 & 780092 & 3.0 & -0.1 & -0.3 \\
$6p^2$ & $^3\!P_1$   &          &1.499 &1.500 &1.501 &  7093  &  7697  &  7710  & 7819   & 9.3 &  1.6 &  1.4\\
$6p^2$ & $^3\!P_2$   &          &1.277 &1.500 &1.269 &  9913  & 10585  & 10587  & 10650  & 6.9 &  0.6 &  0.6\\
$6p^2$ & $^1\!D_2$   &          &1.223 &1.000 &1.230 & 19965  & 21401  & 21440  & 21458  & 7.0 &  0.3 &  0.1\\
$6p^2$ & $^1\!S_0$   &          &      &      &      & 28084  & 29707  & 29808  & 29467  & 4.7 & -0.8 & -1.2\\[0.2pc]
$6p7p$ & $^3\!P_1$   &$^3\!D_1$ &0.671 &0.500 &      & 40732  & 42528  & 42755  & 42919  & 5.1 &  0.9 &  0.4\\
$6p7p$ & $^3\!P_0$   &          &      &      &      & 42303  & 44059  & 44299  & 44401  & 4.7 &  0.8 &  0.2\\
$6p7p$ & $^3\!D_1$   &$^3\!P_1$ &1.469 &1.500 &      & 42486  & 44299  & 44522  & 44675  & 4.9 &  0.8 &  0.3\\
$6p7p$ & $^3\!D_2$   &          &1.173 &1.167 &      & 42630  & 44438  & 44657  & 44809  & 4.9 &  0.8 &  0.3\\[0.2pc]

$6p7s$ & $^3\!P^o_0$ &          &      &      &      & 33104  & 34634  & 34917  & 34960  & 5.3 &  0.9 &  0.1\\
$6p7s$ & $^3\!P^o_1$ &          &1.350 &1.500 &1.349 & 33451  & 34959  & 35243  & 35287  & 5.2 &  0.9 &  0.1\\[0.2pc]
$6p6d$ & $^3\!F^o_2$ &          &0.790 &0.667 &0.796 & 43818  & 45660  & 45933  & 45443  & 3.6 & -0.5 & -1.1\\
$6p6d$ & $^3\!D^o_2$ &          &1.254 &1.167 &1.247 & 44631  & 46458  & 46756  & 46061  & 3.1 & -0.9 & -1.5\\
$6p6d$ & $^3\!D^o_1$ &          &0.883 &0.500 &0.864 & 44714  & 46515  & 46820  & 46068  & 2.9 & -1.0 & -1.6\\
$6p6d$ & $^3\!F^o_3$ &          &1.122 &1.083 &1.116 & 45187  & 46824  & 47134  & 46329  & 2.5 & -1.1 & -1.7\\[0.2pc]
$6p7s$ & $^3\!P^o_2$ &          &1.486 &1.500 &1.496 & 45629  & 47959  & 48282  & 48189  & 5.3 &  0.5 & -0.2\\
\end{tabular}
\end{ruledtabular}
\end{table*}

To illustrate the role of core-valence correlations we calculated
the low-lying energy levels using three different approaches of increasing
accuracy: (i) using the conventional
CI method, (ii) in the framework of the approach combining CI with the
second order of many-body perturbation theory (CI+MBPT method
\cite{DzuFlaKoz96}), and (iii) using the CI method combined with
linearized coupled cluster single-double method (CI+all-order
approach \cite{SafKozJoh09}) modified as discussed in
Sec.~\ref{ManyAt}.

Calculations at the CI+MBPT and CI+all-order stages require
knowledge of matrix elements of the operator $\Sigma$. 
We emphasize that for the $V^{N-2}$ approximation $N_{DF} > N_c$ and $U \neq 0$ and
the modified equations~\eqref{sigma_ang} should be used.
These equations include tilded one-electron
energies $\tilde \varepsilon_v$ of valence orbitals, which still
have to be defined. 
When we are interested only in the low-lying energy levels, an energy dependence of 
the effective Hamiltonian \eqref{Heff1} can be usually neglected for the properly chosen $\tilde \varepsilon_v$.
The recipe of Ref.~\cite{SafKozJoh09} is to put $\tilde \varepsilon_v=
\varepsilon_{v_0}$, where $v_0$ is the lowest valence orbital
for the particular partial wave.
Here we found that the best choice is
\begin{equation}
\tilde \varepsilon_{v} = \varepsilon_{v_0} - U_{{v_0}{v_0}},
\end{equation}
where $U_{{v_0}{v_0}}$ can be obtained from \eref{UDF}. Effectively,
this means that we choose $\tilde \varepsilon_v$ to be the DF energy
of the lowest valence orbital for the given partial wave in the
$V^{N-4}$ potential.

The results of the energy level calculations are presented
in~\tref{EnVN-2}. We find that the accuracy of the CI+MBPT
energies was improved by a factor of 2.5 to 30 in comparison with the CI results
for all energy levels.
We note that a number of energy levels were
reproduced with an accuracy a few tenth percent at the CI+MBPT stage. For such a heavy multivalent
atom as Pb, it looks unexpectedly good and is probably accidental. For this reason further
improvement of an agreement between the theoretical and experimental
energy levels at the CI+all-order stage is difficult.
Both methods underestimate transition energies to the levels of the $6p\,6d$ configuration,
but the results obtained at the CI+MBPT stage are slightly
closer to the experimental values. For almost all other energy levels the
CI+all-order approach gives better agreement with the
experiment, with the average difference with experiment  being 0.6\%.

Our calculation of $g$ factors for the low-lying states revealed a
discrepancy with the NIST database~\cite{RalKraRea11} for two
electronic terms. In the second column we present the electronic
terms provided by NIST~\cite{RalKraRea11}. In the third column we
give our assignment when it differs from the NIST terms. In the columns 4-6 we
present $g$ factors obtained in our calculation, the values corresponding to
the $LS$-coupling scheme, and the experimental numbers. We see rather good agreement between theory and
experiment for all cases where experimental $g$ factors are known. For the $6p7p$ 
configuration, the experimental $g$ factors are unknown.
Calculated $g$ factors indicate some mixing between $LS$ terms, and
support new assignments.
We note that for the less than half filled $p$ shell one should expect
``normal'' order of levels of the $^3\!P_J$ triplet, when the levels with
smaller $J$ are lying lower (see, e.g., the book of \citet{Sob79}).
New term assignments are in agreement with this rule.
\subsection{Hyperfine structure constants}
\label{A:hfs}
Our goal is to calculate the $E1$ PNC amplitude, which is sensitive
to behavior of the wave functions at the nucleus. To test the
quality of the wave functions in the vicinity of the nucleus, we
carried out calculation of the magnetic dipole hyperfine structure
constants $A$ for the even- and odd-parity low-lying states.
We calculate the $E1$ PNC amplitude for the zero spin isotope
$^{208}$Pb, which was used in the
experiments~\cite{MeeVetMaj93,EdwPhiBai95}. Our results for the HFS
constants correspond to the $^{207}$Pb isotope that has nuclear spin
$I=1/2$ and the magnetic moment $\mu/\mu_N \approx
0.5783$~\cite{BouGouHan01}, where $\mu_N$ is the nuclear magneton.
\begin{table*}[htb]
\caption{$V^{N-2}$ approximation. The breakdown of different contributions to the magnetic dipole HFS constants $A$ (in MHz).
The CI+MBPT and CI+all-order values are presented in third and fourth columns, correspondingly. The remaining columns give various
corrections described in the text. Values labeled ``Total'' are obtained as $A$(CI+All)+RPA+Sbt+$\sigma$+SR+Norm.
The recommended values, labeled as ``Recomm.'', are obtained as $A$(CI+All)+RPA+Sbt+$\sigma$+(1/2)SR+Norm (see an explanation in the text).
Last three columns are the experimental results available in the literature.}
\label{VN-2:hfs}
\begin{ruledtabular}
\begin{tabular}{rccrrrrrrrrccc}
                    & CI   &CI+MBPT& CI+All& RPA & Sbt &$\sigma$&  SR &Norm & Total & Recomm.& Refs.~\cite{WasDroKwe05,WasDroKwe07}
                                                                                                       & Ref.~\cite{LurLan70}
                                                                                                                     &  Ref.~\cite{BouGouHan01}\\
\hline \\
$6p^2\,\,^3\!P_1$   &-2184 & -2545 & -2513 &  46 & -28 &  118  & -116 &  45 & -2449 & -2392 &-2416(36) &             & -2389.4(0.7) \\
$6p^2\,\,^3\!P_2$   & 2067 &  2335 &  2369 & 341 &   9 &  -96  & -122 & -48 &  2453 &  2513 & 2739(10) &             &  2600.8(0.9) \\
$6p^2\,\,^1\!D_2$   &  481 &   499 &   519 & 139 &  -2 &   -7  & -121 & -11 &   518 &   577 &  620(6)  & 609.820(8)  &      \\[0.2pc]
$6p7p\,\,^3\!P_1$   & 5914 &  6635 &  6649 & 434 & -24 & -271  &  -83 & -92 &  6614 &  6654 &          &             &   \\
$6p7p\,\,^3\!D_1$   &-2536 & -2886 & -2888 &-181 &  11 &  122  &   27 &  39 & -2868 & -2882 &          &             &   \\
$6p7p\,\,^3\!D_2$   & 2811 &  3144 &  3154 & 227 &  -9 & -128  &  -47 & -44 &  3153 &  3176 &          &             &   \\[0.2pc]

$6p7s\,\,^3\!P_1^o$ & 7785 &  8536 &  8528 & 632 & 121 & -329  &  -74 &-123 &  8753 &  8790 & 8819(14) &             &  8802.0(1.6)  \\
$6p6d\,\,^3\!F_2^o$ & 2633 &  2989 &  2998 & 205 &  -5 & -120  &  -42 & -45 &  2990 &  3011 & 3094(9)  &             &   \\
$6p6d\,\,^3\!D_2^o$ & -827 & -1482 & -1482 & -94 &  13 &   65  &   17 &  33 & -1448 & -1456 &          &             &   \\
$6p6d\,\,^3\!D_1^o$ &-2462 & -2808 & -2816 &-156 & -68 &  120  &  -10 &  41 & -2889 & -2884 &          &             &   \\
$6p6d\,\,^3\!F_3^o$ & 1779 &  1993 &  2000 & 140 &  81 &  -82  &  -27 & -44 &  2066 &  2079 & 2072(8)  &             &   \\[0.2pc]
$6p7s\,\,^3\!P_2^o$ & 1336 &  1593 &  1604 & 287 & -40 &  -43  &  -94 & -26 &  1688 &  1734 &          &             &   \\
\end{tabular}
\end{ruledtabular}
\end{table*}

For an accurate calculation of the HFS constants we take into
account not only random-phase approximation (RPA) corrections but
also the corrections beyond RPA, including one- and two-particle
subtraction contributions (their sum is labeled as ``Sbt''), the
core-Brueckner ($\sigma$), structural radiation (SR), and
normalization (Norm) corrections~\cite{DzuKozPor98}. The results of
the calculation are presented in~\tref{VN-2:hfs}.

The values in the column labeled ``Total'' were found as the sum of
the values obtained at the CI+all-order stage plus the corrections
listed in~\tref{VN-2:hfs}, i.e., $A$(Total) =
$A$(CI+All)+RPA+Sbt+$\sigma$+SR+Norm. We find that
 the corrections (beynod RPA) are sufficiently
large as demonstrated in \tref{VN-2:hfs}. In particular, they are very significant  for the even-parity
states belonging to the $6p^2$ configuration. For example,  the absolute value of
the RPA correction is 2.5 times smaller than the SR correction for the $6p^2\,\,^3\!P_1$ state. We can  explain it as
follows. The main configuration, contributing 94\% in probability to
this state, is $6p_{1/2}\,6p_{3/2}$. Single-electron contributions
of the $6p_{1/2}$ and $6p_{3/2}$ electrons to the HFS constant
$A(6p^2\,\,^3\!P_1)$, are such that they tend to cancel each other.
This holds for the ``bare'' $H_{\rm hfs}$ operator and when we
include the RPA corrections.
As a result, the total RPA correction is not large. The SR
corrections to the single-electron contributions of the $6p_{1/2}$
and $6p_{3/2}$ electrons, in contrast, are added, 
resulting in a large contribution to the HFS constant.

It is worth noting that we calculate the SR corrections only in the
2nd order of the MBPT. Usually the 2nd order of the MBPT
overestimates the respective contribution. We assume that an
inclusion of higher orders (beyond second order) will reduce the absolute value of the SR
contribution. Our values in the column labeled
``Recomm.'' were obtained as described above, with  the SR
corrections reduced by a factor of  two, i.e., $A$(Recom.) =
$A$(CI+All)+RPA+Sbt+$\sigma$+(1/2)SR+Norm. The difference between
calculated and recommended values does not exceed 4\% except for the
level $6p^2\,^1\!D_2$.
The $^1\!D_2$ HFS constant is a few times smaller than other, but the SR 
correction is of comparable size, contributing at the level of 20\%. It leads to a
slightly larger difference (5.4\%) between our value and the most accurate
experimental result~\cite{LurLan70}.

Our recommended values for the HFS constants show  better agreement
with the experimental results \cite{BouGouHan01,LurLan70}. Note that the
experimental values~\cite{WasDroKwe05,WasDroKwe07} are less
accurate. Moreover, their value for the $6p^2\,^3\!P_2$ level disagrees both with the experiment \cite{BouGouHan01} and
with our calculation. Therefore, we do not rely on this experimental
result in estimating the accuracy of our value for the HFS $6p^2\,^3\!P_2$ constant.
Using remaining experimental data and considering the difference between the CI+MBPT
and CI+all-order results, as well as the size of the SR correction,
we estimate the theoretical uncertainties of the HFS constants to be at the level of 4\%.
\subsection{$E1$ transition amplitudes and polarizability}
\label{E1}
The expression for the $E1$ PNC amplitude (in the 2nd
order of the perturbation theory) involves also the matrix elements
of the electric dipole operator. As a result, it is sensitive
to the behavior of the wave functions at long distances.
To test it we calculated a number of $E1$ transition amplitudes relevant to the
$E1$ PNC amplitude of the $6p^2\, ^3\!P_0 \rightarrow 6p^2\,
^3\!P_1$ transition. We  also calculated the value of the $6p^2\,^3\!P_0$ ground state
static polarizability.

For the E1 matrix elements, all corrections beyond RPA (in contrast with the
HFS constants) are relatively small and we present only the final
values of a few most important matrix elements. These values are
obtained in the same way as above: $D$(Total) =
$D$(CI+All)+RPA+Sbt+$\sigma$+SR+Norm, where $D \equiv |\langle
\gamma' ||d|| \gamma \rangle|$ and $\bf d = -r$ is the electric
dipole operator.

The calculated MEs are presented in~\tref{VN-2:E1} and compared with
the values extracted from the experimental transition probabilities.
Unfortunately, the accuracy of the available experimental data is
not very high. For example, the difference between results of
\cite{AloColHer01} and~\cite{Alo96} for the $\langle 6p^2\,\,^3\!P_0 ||d|| 6p7s\,\,^3\!P_1^o \rangle$ matrix element
is about 12\%. For two transitions our calculated
values agree with the experiment, taking into account their error bars. However,
for the transition $6p^2\,\,^3\!P_0 - 6p6d\,\,^3\!D_1^o$ our result
differs from the experiment by 20\%. We do not see an
obvious reason for this discrepancy.

To further test the accuracy of the $E1$ transition amplitudes from the ground state $6p^2\,^3\!P_0$, we calculated its
static polarizability.
Our value, 46.5 a.u., is in a very good agreement with the central value of the experimental result, $47(7)$ a.u.~\cite{TASS08}.

\begin{table}[htb]
\caption{$V^{N-2}$ approximation. The reduced matrix elements
$|\langle f ||d|| i \rangle|$ (in a.u.) for the electric-dipole transitions, obtained
in the CI+all-order approximation  and including RPA, Sbt, $\sigma$, SR, and
normalization corrections. In last column the matrix elements
extracted from the experimental transition probabilities are
presented. The value of the ground state static polarizability
is given in the last line.}
\label{VN-2:E1}
\begin{ruledtabular}
\begin{tabular}{lcc}
   \multicolumn{1}{l}{Transition}     &  \multicolumn{1}{c}{This work}
                                                & \multicolumn{1}{c}{Experim.} \\
\hline \\[-0.5pc]
$6p^2\,\,^3\!P_1 - 6p7s\,\,^3\!P_0^o$ &  1.89   & 2.04(7)$^{\rm a}$ \\
                                      &         & 2.05(10)$^{\rm b}$ \\
%
$6p^2\,\,^3\!P_0 - 6p7s\,\,^3\!P_1^o$ &  1.32   & 1.37(4)$^{\rm a}$ \\
                                      &         & 1.20(5)$^{\rm c}$ \\
$6p^2\,\,^3\!P_0 - 6p6d\,\,^3\!D_1^o$ &  2.01   & 1.62(4)$^{\rm a}$ \\
                                      &         & 1.67(8)$^{\rm b}$ \\[1mm]
$\alpha(6p^2\,\,^3\!P_0)$             &  46.5   & 47(7)$^{\rm d}$
\end{tabular}
\end{ruledtabular}
 $^{\rm a}$Ref.~\cite{AloColHer01};
 $^{\rm b}$Ref.~\cite{PenSla63};
 $^{\rm c}$Ref.~\cite{Alo96};
 $^{\rm d}$Ref.~\cite{TASS08}.
\end{table}

\subsection{PNC amplitude}
\label{PNC}
The parity-nonconserving nuclear spin-independent part of
electron-nuclear interaction can be written as follows:
\begin{eqnarray}
H_{\rm PNC} = -\frac{G_F}{2 \sqrt{2}}\, Q_W \gamma_5 \rho({\bm r}),
\label{HPNC}
\end{eqnarray}
where  $G_F \approx 2.2225 \times 10^{-14}$ a.u. is the Fermi
constant of the weak interaction, $Q_W$ is the nuclear weak charge
given by~\eref{QW}, $\gamma_5$ is the Dirac matrix, and $\rho({\bm r})$ is the nuclear density distribution.

We assume that the nucleus is a uniformly charged ball:
$$\rho({\bm r}) = \frac{3}{4 \pi R^3} \Theta (R-r),$$
where $\Theta (R-r)$ is the Heaviside step function.
The root-mean-square (rms) charge radius for $^{208}$Pb was measured to
be $R_{\rm rms}$ = 5.5010 fm~\cite{Ang04}.  Using the formula
$ R = \sqrt{5/3}\, R_{\rm rms},$
we find $R \approx 7.1108$ fm.

If $|i \rangle$ and $|f \rangle$ are the initial and final atomic
states of the same nominal parity then, to the lowest nonvanishing
order, the electric dipole transition ME is equal to:
\begin{eqnarray}
   \langle f | d_{q,\rm PNC}  | i \rangle  &=&  \sum_{n} \left[
\frac{\langle f | d_q | n  \rangle
      \langle n | H_{\rm PNC} | i \rangle}{E_i - E_n}\right. \nonumber \\
&+& \left.\frac{\langle f | H_{\rm PNC} | n  \rangle
      \langle n | d_q | i \rangle}{E_f - E_n} \right],
\label{e2}
\end{eqnarray}
where $E_i$, $E_f$ and $E_n$ are the energies of the initial, final and
intermediate states, respectively, $q=0,\pm1$, and $|a \rangle \equiv |J_a,M_a \rangle$ with $J_a$ and
$M_a$ being the total angular momentum and its projection.

Taking into account that $H_{\rm PNC}$ is a pseudo-scalar operator, i.e., its ME is nonzero only for the states with
the same $J$ and $M$, we can determine the spin-independent PNC amplitude of
the $6p^2\,\,^3\!P_0 \rightarrow 6p^2\,\,^3\!P_1$ transition, $E1_{\rm PNC}$, as the reduced ME of the electric dipole moment
operator $d_{q,\rm PNC}$:
\begin{eqnarray}
E1_{\rm PNC} &\equiv& \langle f || d_{\rm PNC} || i \rangle \nonumber \\
&=&   \sum_n \left(
\frac{\langle ^3\!P_1||d||n\rangle \langle n|H_{\rm PNC}|^3\!P_0 \rangle} {E_{^3\!P_0}-E_n} \right. \nonumber \\
 &&+ \left. \frac{\langle ^3\!P_1|H_{\rm PNC}|n \rangle \langle n||d||^3\!P_0\rangle} {E_{^3\!P_1}-E_n} \right) \nonumber \\
&\equiv& E1^{(1)}_{\rm PNC} + E1^{(2)}_{\rm PNC}.
\label{E1PNC}
\end{eqnarray}
Introducing notations
\begin{subequations}\label{psi}
\begin{eqnarray}
|\delta \psi_1 \rangle  &=& \sum_n \frac{|n\rangle \langle n|H_{\rm PNC}|^3\!P_0 \rangle}{E_{^3\!P_0}-E_n}, \\
\langle \delta \psi_2 | &=& \sum_n \frac{\langle ^3\!P_1|H_{\rm PNC}|n \rangle \langle n|}{E_{^3\!P_1}-E_n},
\end{eqnarray}
\end{subequations}
 we express $E1^{(1)}_{\rm PNC}$ and $E1^{(2)}_{\rm PNC}$ as
\begin{subequations}\label{E1PNCa}
\begin{eqnarray}
 E1^{(1)}_{\rm PNC} &=& \langle ^3\!P_1||d||\delta \psi_1 \rangle, \\
 E1^{(2)}_{\rm PNC} &=& \langle \delta \psi_2 ||d||^3\!P_0 \rangle.
\end{eqnarray}
\end{subequations}

The $E1_{\rm PNC}$ amplitude is sensitive to the matrix elements of the weak
interaction $H_{\rm PNC}$, $E1$ transition amplitudes, and the energy spectrum.
The weak interaction depends on the wave function in the
vicinity of the nucleus and, in this respect, is similar to the
matrix elements of the hyperfine interaction. Thus, we are able to
estimate the accuracy of the calculation of the PNC amplitude
analyzing the accuracy of the HFS constants and $E1$ transition amplitudes.

In calculating the PNC amplitude we included the RPA corrections,
one- and two-particle subtraction contributions,
the core-Brueckner, structural radiation, and normalization
corrections, as we did when calculated the HFS constants.

When the $E1_{\rm PNC}$ transition amplitude is obtained, we are able to find the quantity
\begin{equation}
R = \frac{{\rm Im}(E1_{\rm PNC})}{M1},
\label{Eq:R}
\end{equation}
where we take into account that $E1_{\rm PNC}$ is imaginary and
designate the reduced matrix element of the magnetic dipole
operator $\mu$: $M1 \equiv \langle 6p^2\,\,^3\!P_1 ||\mu||
6p^2\,\,^3\!P_0 \rangle$. The quantity $R$ was experimentally
determined in~\cite{MeeVetMaj93,MeeVetMaj95,EdwPhiBai95}, so we are
able to compare theory and experiment.

The results of calculation of both $E1^{(1)}_{\rm PNC}$ and
$E1^{(2)}_{\rm PNC}$ terms, determined by Eq.~\eqref{E1PNC}, are
presented in~\tref{VN-2:PNC}.
\begin{table}
\caption{$V^{N-2}$ approximation. The breakdown of different
contributions to the terms $E1^{(1)}_{\rm PNC}$ and $E1^{(2)}_{\rm
PNC}$ determined by~\eref{E1PNC} (in a.u.). The values of $M1\equiv
\langle 6p^2\,\,^3\!P_1 ||\mu|| 6p^2\,\,^3\!P_0 \rangle$ are in the
Bohr magnetons. The values of $R$ are given in units $10^{-8}
\cdot (-Q_W/N)$. First, second, and third lines give the CI,
CI+MBPT, and CI+all-order values, respectively. The following
lines give various corrections described in the text. Numbers
labeled ``Total'' are obtained as (CI+All)+RPA+$\sigma$+SR+Sbt+Norm.
Numbers labeled ``Recomm.'' are obtained as
(CI+All)+RPA+$\sigma$+(1/2)SR+Sbt+Norm (see the explanation in the text).} 
\label{VN-2:PNC}
\begin{ruledtabular}
\begin{tabular}{lrrrr}
         & $E1^{(1)}_{\rm PNC}$ & $E1^{(2)}_{\rm PNC}$ & $M1$  & $R$   \\
\hline \\
CI       &          2.619       &         2.109        &-1.297 & -9.99 \\
CI+MBPT  &          2.768       &         2.495        &-1.292 &-11.16 \\
CI+All   &          2.718       &         2.488        &-1.293 &-11.03 \\
RPA      &          0.344       &        -0.312        &       & -0.07 \\
$\sigma$ &         -0.099       &        -0.077        &       &  0.37 \\
SR       &         -0.032       &         0.086        &       & -0.11 \\
Sbt      &         -0.007       &         0.021        &       & -0.03 \\
Norm     &         -0.055       &        -0.042        &       &  0.21 \\
Total    &          2.869       &         2.164        & 1.293 &-10.66 \\[0.2pc]
Recomm.  &          2.885       &         2.121        & 1.293 &-10.6(4) \\[0.2pc]
Other    &                      &                      &       &-10.4(8)$^{\rm a}$ \\
         &                      &                      &       &-11.4$^{\rm b}$ \\
         &                      &                      &       &-13$^{\rm c}$
\end{tabular}
$^{a}$Ref.~\cite{DzuFlaSil87E}; $^{b}$Ref.~\cite{BotBlaMar87}; $^{c}$Ref.~\cite{NovSusKhr76}.
\end{ruledtabular}
\end{table}
Our analysis shows that the intermediate state $6p7s\, ^3\!P^o_0$ gives
dominating ($\sim$86\%) contribution to $E1^{(1)}_{\rm PNC}$.
Thus, the contribution of higher-lying states is rather small.

For the $E1^{(2)}_{\rm PNC}$ part of the $E1_{\rm PNC}$ amplitude
the situation is quite different. Two lowest-lying odd-parity
states with $J=1$ listed in~\tref{EnVN-2} contribute to
$E1^{(2)}_{\rm PNC}$ with different signs and their total
contribution is negative, i.e., it has a different sign in comparison
with the total value of $E1^{(2)}_{\rm PNC}$. As a result,
higher-lying states give very large contribution to this amplitude.

Such anomalously large contribution  
comes from the high-lying odd-parity states belonging to the configuration $6s\, 6p^3$.
According to our calculation the lowest state with $J=1$, belonging to 
this configuration, is lying  $\sim\! 74000$ cm$^{-1}$
above the ground state. The matrix elements of the electric dipole and $H_{\rm PNC}$ operators are large:
$\langle 6s^2 6p^2 \,\, ^3\!P_0 ||D|| 6s 6p^3 \, J=1 \rangle$ = 1.91 a.u. and
$\langle 6s 6p^3 \, J=1 |H_{\rm PNC}| 6s^2 6p^2 \,\, ^3\!P_1 \rangle$ = 476 a.u..
As a result, the contribution of this odd-parity state to $E1^{(2)}_{\rm PNC}$ is large and positive.

It is worth noting that, for the reason discussed above, a direct
summation over intermediate states is not applicable for calculation
of $E1^{(2)}_{\rm PNC}$. Instead, we solve inhomogeneous equation \cite{KPF96} which accounts for contribution 
from all discrete states and a continuum.

Analyzing the RPA and other corrections to $E1^{(1)}_{\rm PNC}$ and
$E1^{(2)}_{\rm PNC}$, we see that large RPA corrections have
different signs for these two amplitudes.
Accidentally, these contributions turned out to be close in their
absolute values and essentially cancel each other in the sum
$E1^{(1)}_{\rm PNC} + E1^{(2)}_{\rm PNC}$. For this reason the role
of smaller corrections ($\sigma$, Sbt, e.t.c) is enhanced.

A procedure of including the RPA, $\sigma$, and SR corrections in
calculating $E1_{\rm PNC}$ is reduced to ``dressing'' the
$H_{\rm PNC}$ and ${\bf d}$ operators, as described in detail in~\cite{DzuKozPor98}.
To find the subtraction and normalization corrections, following the recipe of Ref.~\cite{DzuKozPor98}, we
obtained $|\delta \psi_1 \rangle$ and $|\delta \psi_2 \rangle$,
given by~\eref{psi}, for the effective operator $H^{\rm eff}_{\rm
PNC}$ and then calculated the MEs in \eqref{E1PNCa}
for the effective electric dipole operator ${\bf d}^{\rm eff}$.

The values listed in the row labeled ``Total'' of~\tref{VN-2:PNC}
were obtained as the sum of the CI+All values plus different
corrections including RPA, $\sigma$, SR, Sbt, and Norm. As we
 discussed above, the SR corrections turn out to be
overestimated in the 2-nd order of the MBPT. We had reduced these
corrections by a factor of 2  to obtain the recommended values of the
HFS constants. We assume that the same procedure should be used for
the PNC amplitude as well, though in this case the SR corrections
are not so significant as for the HFS constants. The results listed
in the row labeled ``Recomm.'' are obtained as the ``Total'' values
but we add only a half of the SR correction. According to our
estimate, the excitations of the core electrons contribute to
$E1_{\rm PNC}$ less than 0.1\% and we neglect this contribution.

The RPA and other similar corrections are
very small for the $M1$ matrix element $\langle 6p^2\,\,^3\!P_1 ||\mu||
6p^2\,\,^3\!P_0 \rangle$ and can be neglected without loss of accuracy. We
present the values of the quantity $R$ (given by~\eref{Eq:R})
obtained in different approximations in the last column of the table. Various corrections to $R$
are listed as well. Our
recommended value is $R = -10.6(4) \times 10^{-8} \, (-Q_W/N)$.
Based on the calculation accuracy of the HFS constants, $E1$
transition amplitudes, and the ground state polarizability, we assign
to the quantity $R$ the uncertainty $\sim 4\%$. Our result is in a
good agreement with earlier calculations
\cite{NovSusKhr76,BotBlaMar87,DzuFlaSil87E} but the accuracy is two
times higher.

Using our recommended value of $R$ and the most accurate
experimental value $(-9.86 \pm 0.12) \times
10^{-8}$~\cite{MeeVetMaj93,MeeVetMaj95} we  find the weak
nuclear charge for $^{208}$Pb to be $Q_W = -117(5)$. This value is
in good agreement with the SM prediction $Q^{\rm SM}_W = -118.79(5)$
\cite{PDG12}. Note that our theoretical error (4\%) is more than
three times larger than the experimental error (1.2\%). Therefore we
need further improvement of the theory for more accurate calculations.
A next step in improving accuracy would be to treat SR corrections to all orders.

\section{$V^{N-4}$ approximation}
\label{VN-4}
It is worth noting that due to importance of the $6s 6p^3$
configuration for the $E1_{\rm PNC}$ amplitude discussed above, a
two-electron calculation of Pb using $V^{N-2}$ approximation gives
poor results. In this section, as earlier, we consider Pb as the
4-valence atom using $V^{N-4}$ approximation and compare results
with those obtained previously in the $V^{N-2}$ approximation.
Both calculations are similar, so we focus here only on the points,
where these two approaches differ from each other.

The basis set was constructed using Dirac-Fock-Sturm approach,
but the Dirac-Fock equations were solved for the [$1s^2,...,5d^{10}$] closed core,
i.e., the $6s$ electrons were excluded from the self-consistency procedure.
Next, the 6--8$s$, 6--8$p$, $6d$, and $4f$ orbitals were constructed
in the field of the frozen core $V^{N-4}$ potential. The virtual orbitals were obtained by solving
Dirac-Fock-Sturm equations~\cite{TupVolGla05,TupKozSha10}.
The constructed basis set included, in total, 6 partial waves with the orbitals up to
$32s, 32p, 32d, 30f, 25g$ and $25h$ just as in the $V^{N-4}$ case.

\begin{table}[htb]
\caption{The energy levels (in cm$^{-1}$) obtained using the
CI+all-order method in $V^{N-2}$ and $V^{N-4}$ approximations are
compared with the experiment~\cite{RalKraRea11}. Four-electron
binding energies are given in the first row for the ground state,
energies in other rows are counted from the ground state.
Corresponding relative differences of these two calculations with
the experiment are given in percentages. Electronic terms in the 2nd
column correspond to the new assignment discussed in text.}
\label{Energies}%
\begin{ruledtabular}
\begin{tabular}{rcccccc}
\multicolumn{1}{r}{\multirow{2}{*}{Conf.}} &
\multicolumn{1}{c}{\multirow{2}{*}{Term}} &
\multicolumn{2}{c}{\multirow{1}{*}{CI+All}} &
\multicolumn{1}{l}{\multirow{2}{*}{Exper.}} &
\multicolumn{2}{c}{\multirow{1}{*}Diff.\ (\%)} \\
&
&\multicolumn{1}{r}{$V^{N-4}$} & \multicolumn{1}{r}{$V^{N-2}$} &
&\multicolumn{1}{r}{$V^{N-4}$} & \multicolumn{1}{r}{$V^{N-2}$} \\
\hline \\
$6p^2$ & $^3\!P_0$   & 781122 & 782396 & 780092 & -0.1 & -0.3 \\
$6p^2$ & $^3\!P_1$   &  7576  &  7710  & 7819   &  3.1 &  1.4\\
$6p^2$ & $^3\!P_2$   & 10434  & 10587  & 10650  &  2.0 &  0.6\\
$6p^2$ & $^1\!D_2$   & 21228  & 21440  & 21458  &  1.1 &  0.1\\
$6p^2$ & $^1\!S_0$   & 29779  & 29808  & 29467  & -1.1 & -1.2\\[0.2pc]
$6p7p$ & $^3\!D_1$   & 42384  & 42755  & 42919  &  1.2 &  0.4\\
$6p7p$ & $^3\!P_0$   & 44017  & 44299  & 44401  &  0.9 &  0.2\\
$6p7p$ & $^3\!P_1$   & 44219  & 44522  & 44675  &  1.0 &  0.3\\
$6p7p$ & $^3\!D_2$   & 44364  & 44657  & 44809  &  1.0 &  0.3\\[0.2pc]

$6p7s$ & $^3\!P^o_0$ & 34444  & 34917  & 34960  &  0.9 &  0.1\\
$6p7s$ & $^3\!P^o_1$ & 34778  & 35243  & 35287  &  0.9 &  0.1\\[0.2pc]
$6p6d$ & $^3\!F^o_2$ & 46603  & 45933  & 45443  & -0.5 & -1.1\\
$6p6d$ & $^3\!D^o_2$ & 47176  & 46756  & 46061  & -0.9 & -1.5\\
$6p6d$ & $^3\!D^o_1$ & 47052  & 46820  & 46068  & -1.0 & -1.6\\
$6p6d$ & $^3\!F^o_3$ & 47715  & 47134  & 46329  & -1.1 & -1.7\\[0.2pc]
$6p7s$ & $^3\!P^o_2$ & 47884  & 48282  & 48189  &  0.5 & -0.2\\
\end{tabular}
\end{ruledtabular}
\end{table}

We used exactly the same sets of configurations for even- and odd-parity states as for the calculations
in the $V^{N-2}$ approximation discussed in previous sections. 
The CI+MBPT and CI+all-order methods were used as discussed in Refs.~\cite{DzuFlaKoz96}
and~\cite{Koz04,SafKozJoh09}.

In~\tref{Energies} we compare the results obtained using the
CI+all-order methods in the framework of the $V^{N-2}$ and $V^{N-4}$ approximations. We find
that the low-lying levels belonging to the $6p^2$, $6p7p$,
and $6p7s$ configurations were reproduced better in the $V^{N-2}$
approximation. The $V^{N-4}$ approximation gives slightly better
agreement with the experiment only for the states of the $6p6d$
configuration.

We also calculated the HFS constants in the $V^{N-4}$ approximation following the procedure
described in Section~\ref{A:hfs}. There are no subtraction diagrams in this case. Accounting for poor 
initial approximation, we expect that certain corrections to the HFS constants  to
be large. In particular, the normalization corrections are about 6\% for all HFS constants.

\begin{table}[ht]
\caption{The magnetic dipole HFS constants (in MHz) obtained in
$V^{N-2}$ and $V^{N-4}$ approximations are compared with the
experimental values, where available. The recommended values for the
$V^{N-2}$ approximation are listed. Corresponding relative
differences of these two calculations with the experimental results
are given in percentages.}
\label{hfs}%
\begin{ruledtabular}
\begin{tabular}{rccccccc}
\multicolumn{1}{r}{\multirow{2}{*}{Conf.}} &
\multicolumn{1}{c}{\multirow{2}{*}{Term}} &
\multicolumn{2}{c}{\multirow{1}{*}{CI+All}} &
\multicolumn{2}{c}{\multirow{2}{*}{Experiment}} &
\multicolumn{2}{c}{\multirow{1}{*}Diff.\ (\%)} \\
&&\multicolumn{1}{r}{$V^{N-4}$} & \multicolumn{1}{r}{$V^{N-2}$} &
&&\multicolumn{1}{r}{$V^{N-4}$} & \multicolumn{1}{r}{$V^{N-2}$} \\
\hline \\
$6p^2$ & $^3\!P_1$   & -2265  & -2392  & -2389.4(0.7) &\cite{BouGouHan01} & 5.2 & -0.1 \\
$6p^2$ & $^3\!P_2$   &  2187  &  2513  &  2600.8(0.9) &\cite{BouGouHan01} & 16  &  3.4 \\
$6p^2$ & $^1\!D_2$   &   453  &   577  &  609.820(8)  &\cite{LurLan70}    & 26  &  5.4 \\
$6p7p$ & $^3\!D_1$   &  6062  &  6654  &              &                   &     & \\
$6p7p$ & $^3\!P_1$   & -2612  & -2882  &              &                   &     & \\
$6p7p$ & $^3\!D_2$   &  2873  &  3176  &              &                   &     & \\[0.2pc]

$6p7s$ & $^3\!P^o_1$ &  7969  &  8790  &  8802.0(1.6) &\cite{BouGouHan01} & 10  & 0.14 \\[0.2pc]
$6p6d$ & $^3\!F^o_2$ &  2678  &  3011  &  3094(9)     &\cite{WasDroKwe05} & 13  & 2.7  \\
$6p6d$ & $^3\!D^o_2$ &  -381  & -1456  &              &                   &     &   \\
$6p6d$ & $^3\!D^o_1$ & -2388  & -2884  &              &                   &     &   \\
$6p6d$ & $^3\!F^o_3$ &  1829  &  2079  &  2072(8)     &\cite{WasDroKwe05} & 12  & -0.4   \\[0.2pc]
$6p7s$ & $^3\!P^o_2$ &   715  &  1734  &              &                   &     &    \\
\end{tabular}
\end{ruledtabular}
\end{table}

A comparison of the HFS constants obtained in $V^{N-2}$ and
$V^{N-4}$ approximations using the CI+all-order method and including the
RPA and other corrections, mentioned in Section~\ref{A:hfs}, is given
in~\tref{hfs}. The available experimental values are also presented.
The results obtained in the $V^{N-2}$ approximation
agree with the experiment significantly better. In total, as is seen
from Tables~\ref{Energies} and~\ref{hfs}, the results obtained in
the $V^{N-4}$ approximation are generally less accurate and this
method of calculation is less reliable.

\section{Conclusion}
\label{Concl}
In this paper we have developed and generalized the CI+all-order method for a more
flexible choice of the initial approximation. Previously, using the
CI+all-order method, it was needed to construct basis sets corresponding
to the self-consistent field of the core. Such basis sets are not very good for the systems with
several valence electrons. Here we derived coupled-cluster equations for the
potential which may include (some of) valence electrons and updated
our package of programs. We used this package to calculate atomic
lead as a four electron system in the $V^{N-2}$ approximation. We
studied different properties, including the energy levels, hyperfine
structure constants, $E1$ transition amplitudes, and the ground state
polarizability.

For comparison we also calculated a number of Pb properties in the
$V^{N-4}$ approximation, i.e., using the self-consistent field of the core. Results of this
calculation appeared to be less accurate. We conclude that for such
a heavy and multivalent atom as Pb our new version of the method gives
better accuracy for different observables and is more reliable.

We used this developed variant of the CI+all-order method to calculate the parity nonconserving
transition amplitude $E1_{\rm PNC}\,(6p^2\,\,^3\!P_0 -
6p^2\,\,^3\!P_1)$. The theoretical accuracy for $E1_{\rm PNC}$ was
improved by a factor of two compared to the most accurate previous
calculation~\cite{DzuFlaSil87E}. Using the value obtained for this
amplitude and the experimental
result~\cite{MeeVetMaj93,MeeVetMaj95}, we found the nuclear weak
charge for $^{208}$Pb to be $Q_W = -117(5)$, which agrees with the
SM prediction. Note that our theoretical error (4\%) is sill more
than three times larger than the experimental error (1.2\%).
Therefore, to calculate more accurately different properties of such
a heavy multivalent atom as Pb, we need further improvement of the
theory. A next step in improving accuracy would be to treat SR corrections to all orders. 

The work of S.G.P.\ and M.S.S.\ was supported in part by U.S. NSF
grants No.\ PHY-1404156 and No.\ PHY-1212442; M.G.K.\ and I.I.T.\
were supported in part by RFBR grants No.\ 14-02-00241 and No.\ 15-03-07644.


\end{document}